\documentclass[preprint]{aastex}
\usepackage{amsmath}
\usepackage{graphicx}
\usepackage{url}

\shorttitle{The peculiar solar minimum was predictable}
\shortauthors{Basu et al.}

\begin{document}

\title{Thinning of the Sun's magnetic layer: the peculiar solar minimum could have been predicted}

\author{
Sarbani~Basu\altaffilmark{1}
Anne-Marie Broomhall\altaffilmark{2}
William J. Chaplin\altaffilmark{2}
Yvonne~Elsworth\altaffilmark{2}
}

\altaffiltext{1}{Department of Astronomy, Yale University, PO Box 208101, New Haven, CT 06520-8101;
sarbani.basu@yale.edu}
\altaffiltext{2}{School of Physics and Astronomy, University of
Birmingham, Edgbaston, Birmingham B15 2TT, UK}

\begin{abstract}
The solar magnetic activity cycle causes changes in the Sun on timescales that
are relevant to human lifetimes. The minimum in solar activity that preceded
the current solar cycle (cycle 24) was deeper and quieter than any other recent
minimum.  Using data from the Birmingham Solar-Oscillations Network (BiSON),
we show that the structure of the solar sub-surface layers during the descending
phase of the preceding cycle (cycle 23) was very different from that during
cycle 22.
This leads us to believe that a detailed examination of the data
would have led to the prediction that the cycle-24 minimum would be out of the ordinary.
The behavior of the oscillation frequencies allows us to infer
that changes in the Sun that affected the oscillation frequencies in cycle
23 were localized mainly to layers above about 0.996R$_\odot$, depths shallower than about 3000 km.
In cycle 22, on the other hand, the changes must have also occurred in the deeper-lying
layers.
\end{abstract}

\clearpage
\section{Introduction}

The solar magnetic activity cycle, usually referred to as the solar cycle, causes
changes on the Sun on time scales relevant to human life on earth. The maxima of
the cycles are marked by large numbers of sunspots, as measured by the International
Sunspot Number (ISN). Also prevalent are solar flares and emissions that are a
direct result of areas of strong magnetic fields on the Sun, e.g., radio emission
at a wavelength of 10.7 cm (RF10.7). The effect that these flares have on the
Earth has led to many efforts to try to predict levels of solar activity led by
the Space Weather Prediction Center of the National Oceanic and Atmospheric
Administration.\footnote[3]{http://www.swpc.noaa.gov/SolarCycle/}

The minimum that preceded the current solar cycle (cycle 24) was unusually quiet
and long, and defied all existing  predictions. Most solar activity markers, such as the
RF10.7, were at an all-time low; the polar magnetic fields were weak and the
amount of local flux was small (see e.g. articles in ASPCS vol. 428).
Observations of interplanetary scintillation (IPS) showed solar-wind turbulence
(a quantity related to solar magnetic fields) had been steadily dropping for a
decade before the deep minimum (Janardhan et al. 2011).
Such a low minimum had not been seen since detailed solar-cycle observations
began with the advent of the satellite era. The predictions for this solar minimum,
including the timing for when the Sun would emerge from it, turned out to be
incorrect raising questions about the validity of the prediction techniques (Hathaway 2010).

The activity cycle has its origins in processes occurring in the solar interior.
While the usual solar-cycle related observations deal with phenomena on the solar
surface or beyond, helioseismic data, i.e., data on solar oscillations, provide
the means to probe beneath the solar surface. A number of observational programs 
such as the Global Oscillations Network Group (GONG), the Michelson Doppler Imager
on board the Solar and Heliospheric Observatory (SoHO) and the Birmingham Solar
Oscillations Network (BiSON)  have been collecting data on solar oscillations.
Helioseismologists use the frequencies of solar oscillations to derive the
structure and dynamics of the solar interior (e.g. Christensen-Dalsgaard 2002).

Solar oscillations respond to the changing solar activity. It has been
shown that the frequencies of solar oscillations increase with activity
(Woodard \& Noyes 1985; Elsworth et al. 1990; Libbrecht \& Woodard 1990). The frequency shifts are
predominantly a function of frequency alone, and all the degree dependence can be
removed by correcting for mode inertia (mode inertia determines how a mode
responds to a given perturbation, high-frequency and high-degree modes have lower
inertia than low-frequency or low-degree modes and hence their frequencies
are changed more easily).
The nature of these solar-cycle related variations is such that most of the
changes to the structure of the Sun are believed to be restricted predominantly to a thin
sub-surface layer (Nishizawa \& Shibahashi 1995). Changes to solar structure in the
deep interior are small and the inferred change in the sound-speed at the base
of the convection zone is only of the order of $10^{-5}$ (Baldner \& Basu 2008).

Dynamical variations do however penetrate the sub-surface convection zone more
deeply, and analysis of GONG and MDI data has shown that a pole-ward flow of
material seen in the solar interior at the beginning of cycle 23 was not seen at
the beginning of cycle 24 (Howe et al. 2009; Antia \& Basu 2010).
Some theoretical studies have suggested that differences in behavior between solar
cycles are caused by variations of flows, particularly the meridional flow,
inside the Sun (Nandy et al. 2011).

Neither the GONG nor MDI observations could predict that cycle 23 would end
in a deep minimum, since there were no data from prior cycles to compare against. However,
BiSON has now been collecting solar oscillations data for almost three complete
solar cycles stretching back over 30 years. In this work we show that an early
comparison of the oscillation frequencies from the descending phases of cycles
22 and 23 would have revealed that we were headed towards a peculiar minimum.
Indeed, as we show below, we would have known that the entire descending phase
of cycle 23 was strange compared to that of cycle 22.

\section{The Data}

BiSON makes Sun-as-a-star (unresolved) Doppler
velocity observations, which are sensitive to the p modes with the
largest horizontal scales (or the lowest angular degree $\ell$), in particular $\ell=$ 0,1,
and 2.
Consequently, the frequencies measured by BiSON are of the truly
global modes of the Sun.  These modes travel to the Sun's core,
however, their dwell time at the surface is longer than at the solar
core because the sound speed inside the Sun increases with depth.
Therefore, the low-$\ell$ modes are also very sensitive to variations in
regions of the interior that are close to the surface and so are
able to give a global picture of the influence of near-surface
activity.
BiSON is in a unique position to study the changes in oscillation frequencies that
accompany the solar cycle as it has now been collecting data for over 30yrs.
{BiSON observations thus cover most of cycle 21 { (coverage beins 1978 July 31; the end of the cycle
is around 1986 April 14)}, the complete cycle 22 (1986 April 14 to 1996 April 11),
and the complete cycle 23 (1996 April 11 to 2008 October 7) and also cycle 24 (2008 October 7 to present).}
To obtain solar cycle related frequency shifts the BiSON data however, must be split into
shorter subsets.

\begin{figure}
\epsscale{0.75}
\plotone{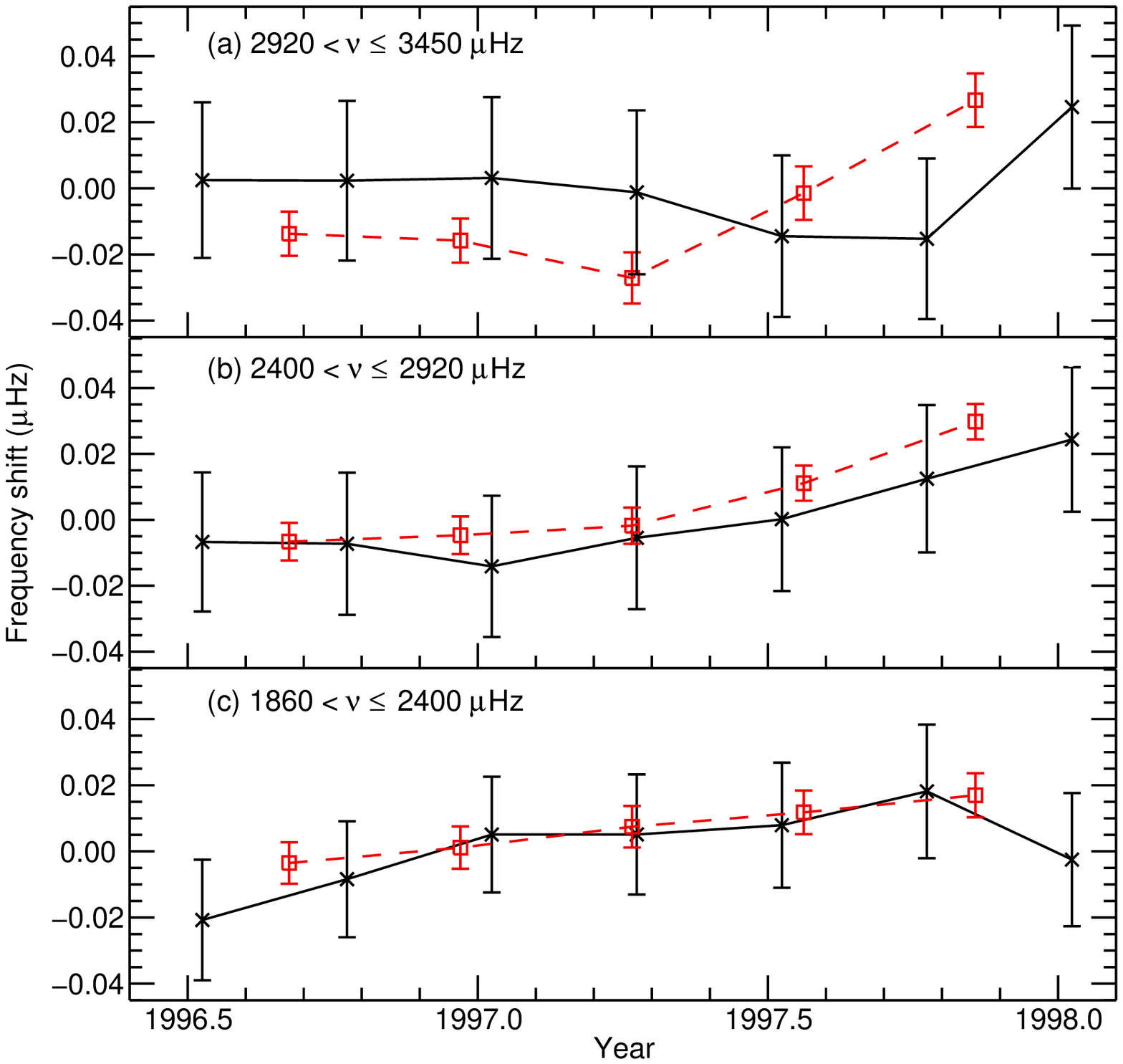}
\caption{{Average frequency shifts of solar oscillations observed by BiSON (black) and GONG (red), as a
function of time { for three different frequency ranges}.
The plotted shifts were calculated with respect to the averaged frequencies of the entire plotted epoch.
The BiSON data-points  are the shifts for $\ell=0$-2 modes. The
GONG points are the shifts for $\ell=8$-10  modes corrected
for the effects of mode inertia.}}
\label{fig:gong}
\end{figure}

As noted above, BiSON does not observe intermediate- or high-degree modes that, along with
low-degree modes help build of  complete image of the Sun and change therein.
However, being the only project that has been observing the Sun without break for more than three
{ decades}, BiSON data are the only ones that will allow us to compare solar cycle 23 with solar cycle 22. 
The lack of intermediate degree modes is however, only a limitation and not an ultimate liability when
comparing solar-cycle related shifts. Chaplin et al. (2001) had shown {that} the BiSON frequency shifts are
commensurate with the frequency shifts {of the } intermediate- and high-degree modes observed by the
Global Oscillation Network Group (GONG; Hill et al. 1996).
We demonstrate this further in Fig. \ref{fig:gong} which 
shows the GONG shifts for modes of $\ell=8$-10 for a few epochs around the cycle 23 minimum {(1996 June 6 to
1998 February 7)}.
The GONG frequency shifts were scaled to account for the effect of mode
inertia in the manner described in Chaplin et al. (2001). 
We also show the BiSON frequency shifts on the same plot.
Shifts are shown for  three frequency ranges {(a) $1860 < \nu \le 2400\,\mu$Hz, (b) 
$2400 < \nu \le  2920\,\mu$Hz
and (c)  $2920 < \nu \le 3450\,\mu$Hz}. To allow a like-for-like comparison the 
reference sets for the BiSON and GONG frequency shifts were defined as the 
average frequencies observed over the epochs  plotted in Fig. \ref{fig:gong}. 
As can be seen,
frequency shifts for the BiSON low-degree modes are in agreement with those for 
GONG. {There are small disagreements between GONG and BiSON in the highest 
frequency range, but even there the disagreements are within $1\sigma$.}
This tells us that we can use the BiSON low-degree modes to probe the Sun
for epochs where no intermediate-degree mode-frequencies are available.

For this work the BiSON data were divided primarily into
365-day-long subsets that overlapped by 91.25\,d. However, this is only true
for the data observed after 1986 April 14. Before this date the coverage of
the BiSON network was sparse and so we have selected time series of various
lengths such that the duty cycles of the subsets {was never less than 16\%}.
This has allowed us to analyze { 34yr's} of data (from 1978 July 31 to 2012 July 7).

Estimates of the mode frequencies were extracted from each subset by
fitting a modified series of Lorentzian models to the data using a
standard likelihood maximization method, which was applied in the
manner described in \citet{Fletcher2009}. A reference frequency set
was determined by averaging the frequencies in subsets covering the
solar cycle 22 maximum activity epoch, which we defined to occur between 1988 October 12 and 1992 April 12. 
This corresponds to the time period when the frequencies were, on average, at 
their highest and so we therefore expect the majority of frequency shifts to be negative. 
It should be noted that the main results described here are
insensitive to the exact choice of subsets used to make the
reference frequency set. Individual frequency shifts were then defined as the
differences between frequencies given in the reference set and the
frequencies of the corresponding modes observed at different epochs
\citep{Broomhall2009a}.

For each subset in time, three weighted-average frequency shifts
were generated over different frequency intervals, where the weights
were determined by the formal errors on the fitted frequencies:
first, a ``low-frequency'' average shift was determined by averaging the
individual shifts of the $\ell=0$, 1, and 2 modes over four
overtones (covering a frequency range of $1860 < \nu \le 2400\,\mu$Hz); second, a ``mid-frequency''
average shift was computed by averaging over overtones whose
frequencies lie in the interval  $2400 < \nu \le  2920\,\mu$Hz; and third, a
``high-frequency'' average shift was calculated using
overtones with frequencies $2920 < \nu \le 3450\,\mu$Hz. The
lower limit of this frequency range (i.e., $1860\,\rm\mu Hz$) was
determined by how low in frequency it was possible to accurately fit
the data before the modes were no longer prominent above the
background noise. 
To maintain consistency we then ensured that all three
frequency ranges covered the same number of overtones i.e. 4.

{
Solar frequency shifts are known to be correlated with solar activity indices (see e.g.,
Basu \& Antia 2000; Jain \& Bhatnagar 2003), however the correlation is not perfect
on short time scales and depends on the proxy used (Jain et al. 2009).
 We use
two sets of solar-activity related data: (1)  the International Sunspot Number (ISN) available
from the National Geophysical Data Center\footnote{http://www.ngdc.noaa.gov},
(2) the flux of the 10.7cm radio emission also available
from the National Geophysical Data Center.

We use two approaches to study differences between solar cycles 22 and 23: (1)  we
compare the frequency shifts in each cycle to solar activity proxies, and (2) we
compare the frequency changes of the low-frequency and medium-frequency modes during  each cycle with the
corresponding shift in the frequencies of the high-frequency modes.
}

\section{Results and Discussion}

\begin{figure}
\epsscale{0.75}
\plotone{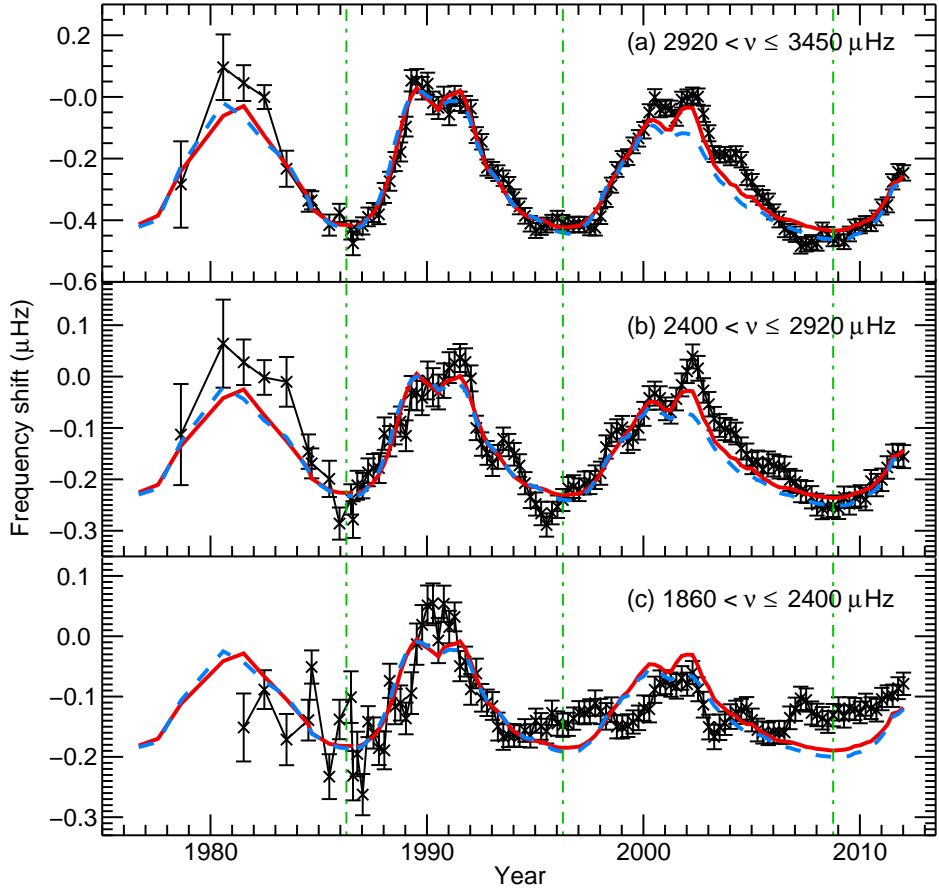}
\caption{Average frequency shifts of solar oscillations observed by BiSON, as a
function of time \textbf{for three different frequency ranges.}
The plotted shifts were calculated with respect to the averaged frequencies during
the cycle 22 maximum, specifically over the period October 1988 to April 1992. 
{ 
Black points in each panel show the average shift in frequency.}
The red line is the 10.7 cm flux, and the blue line is the
International Sunspot Number (ISN), both scaled to match the frequency shifts
observed during the maximum of cycle 22. The vertical lines mark the epochs of
solar activity minimum.}
\label{fig:unsmoo}
\end{figure}

In Fig.~\ref{fig:unsmoo} we show the average shifts in solar oscillation frequencies as a function of time, for the
three different frequency ranges. {Note that some of the frequency shifts around the maximum of cycle 22 are positive because the
reference set that was subtracted out corresponded to an average at the maximum of cycle 22. 
Therefore, occasionally the mode frequencies in an individual 365\,d time series were 
marginally higher than the average set.} However, this by no means implies that the solar frequencies were negative
at subsequent time (which would be unphysical anyway) since the shifts are four orders of magnitude
smaller than the frequencies themselves.

Overplotted on the same figure are RF10.7 and ISN, scaled to match the
frequency shifts observed during cycle 22. { The scaling procedure is described 
below.}
The data for each proxy were averaged over the same epochs in time that were used
to obtain the frequency shifts e.g. after 1986 April 14 the data were
averaged over 365\,d periods that overlapped by 91.25\,d. Before this date the
coverage of the BiSON data were more sparse and so the proxy data used were also
adapted accordingly.  {  The scaling was done by performing a linear least-squares
fit between frequency shifts and the equivalent mean proxy value,
$\bar{F}$ calculated over the period from which each individual frequency shift was
calculated. The scaled value of the proxy, $F_s$, was therefore given by $F_s=m\bar{F}+c$,
where $m$ and $c$ are values obtained from the least-squares fit. The constant term is required
since the proxies are non-zero even when the frequency shifts are zero.
The averaged data were scaled to match the frequency shifts observed during
cycle 22 (during the period 1988 October 12 to 1992 April 12).
This scaling was done separately for all three frequency ranges.}

Note that in cycle 22, the
frequency shifts and the two proxies were in lockstep. Although the BiSON data
for cycle 21 are of poorer quality, it appears that the oscillation frequencies
and the activity proxies behaved in a manner similar to that in cycle 22. However,
in cycle 23 the frequencies and the two proxies deviated from one another. The
deviations became particularly large on the descending phase of the solar
cycle 23, and in the minimum that followed.

{ In addition to the solar-cycle related shift, the frequency shifts shown in Fig.~\ref{fig:unsmoo} illustrate the
two-year cycle first reported by Fletcher et al. (2010) and examined further by Broomhall et al.~(2012)
and Simoniello et al. (2012).
Fletcher et al.~(2010) and Broomhall et al.~(2012) showed that in addition to the signature of the
11 year solar cycle, the signature of a 2-year cycle is visible in the frequency shifts. These
can be seen in Figs.~1 and 2 of Fletcher et al. and Fig.~3 of Broomhall et al. The signature of the
two-year cycle is somewhat different from that of the 11-year cycle. The 11-year
solar cycle causes the frequencies of higher frequency modes to shift more than that
of medium- and low-frequency modes. The frequency shifts due to the 2-year cycle on the
other hand are less dependent on frequency with
relatively more power at lower frequencies pointing to a deeper origin. The amplitude of the cycle however
seems to be modified by the 11-year cycle, and the amplitude is largest at solar maximum and
smallest at the solar minimum. Fletcher et al.~(2010) and Broomhall et al. (2012) have speculated that the
dynamo action causing this phenomenon could be seated near the bottom of the surface layer (about 5\% in depth)
that shows strong rotation shear just like in the tachocline. The presence of two different
dynamos operating at different depths had earlier been proposed by Benevolenskaya (1998a,b)
to explain the quasi-biennial behavior of solar activity proxies.
}

\begin{figure}
\epsscale{0.75}
\plotone{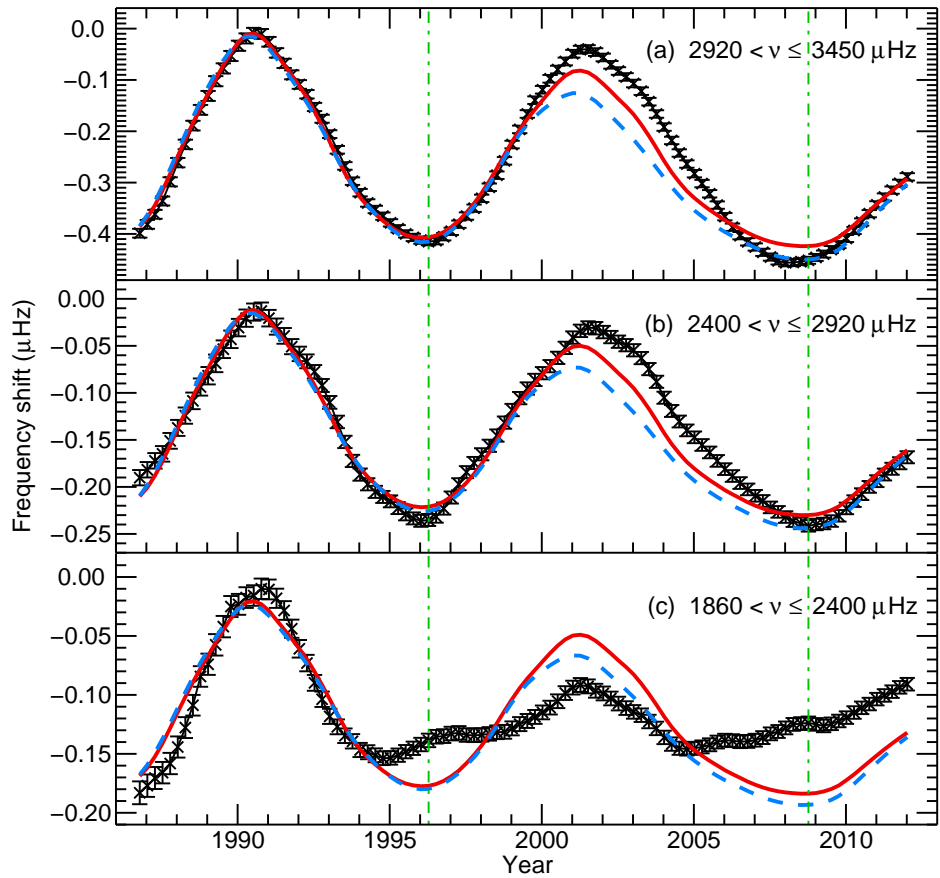}
\caption{The same as Fig.~\ref{fig:unsmoo}, but with the data smoothed to remove the signature
of the two-year cycle. The frequency shifts and the proxies were smoothed
in an identical fashion.}
\label{fig:smoo}
\end{figure}

Since the two-year cycle's signature hides the mismatch between
the proxies and the frequency shifts, we show the same data in Fig.~\ref{fig:smoo}, but with the
signature of the 2-year cycle removed in order to accentuate the roughly 11-year solar cycle related frequency
shifts. Cycle 21 is not plotted in the figure since there
are not enough data to perform a proper smoothing.  The data on the activity proxies
were smoothed in an identical manner. The mismatch between the activity
proxies and the solar frequency shifts is clearer in this plot, and this figure also allows us
to clearly see that the low-frequency shifts began to behave very strangely from the very
onset of cycle 23.

\begin{figure}
\epsscale{0.75}
\plotone{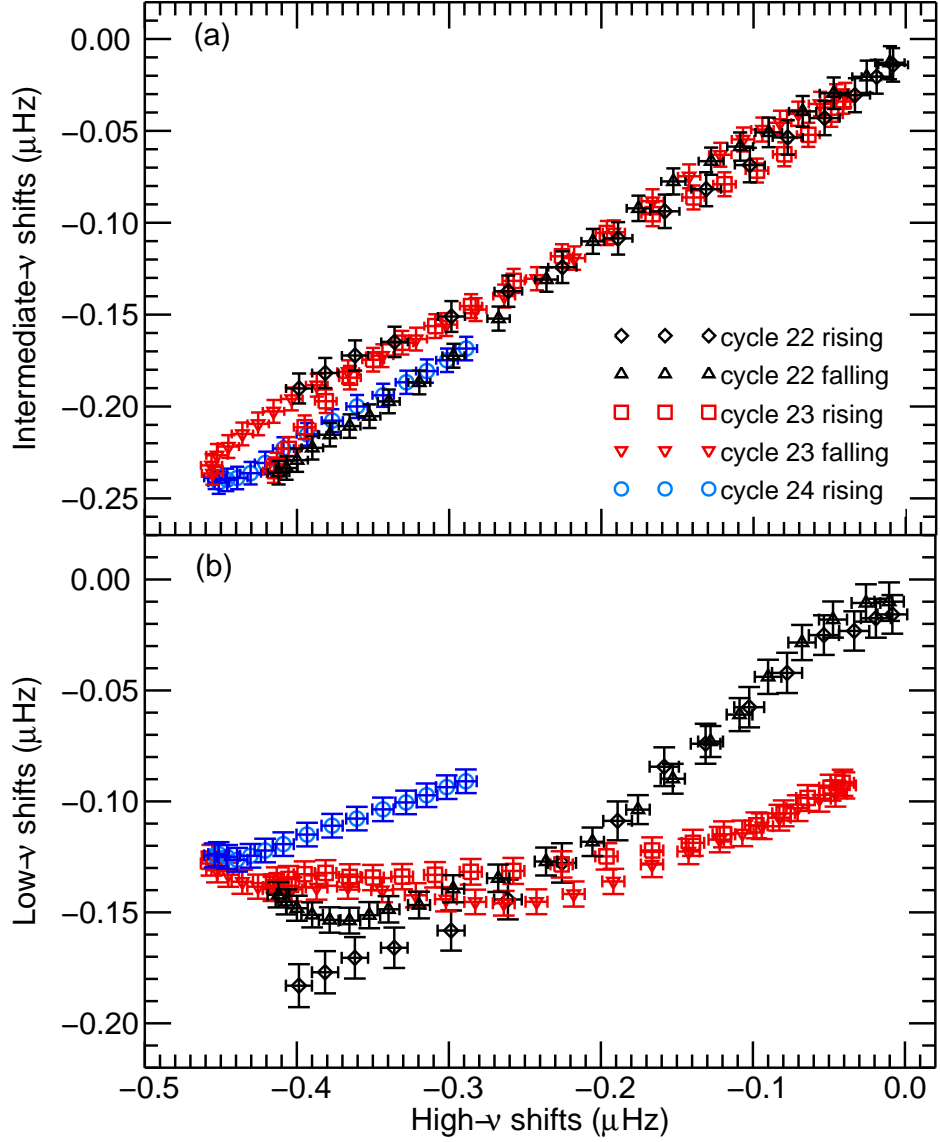}
\caption{
The frequency shifts in the mid-frequency range
(panel a) and low-frequency range (panel b) plotted as a function of the
frequency shifts in the high-frequency range for cycles 22, 23 and 24. The
frequency shifts have been smoothed to remove the signature of the two-year cycle.}
\label{fig:ratio}
\end{figure}

In order to further examine the differences between cycles 22 and 23, and to localize the
differences between the cycles we examine the behavior of the low- and mid-frequency
shifts as a function of the high-frequency shifts for each cycle.  This can be seen in Fig.~\ref{fig:ratio}.
Note that in cycle 22 both the mid- and low-frequency
modes changed in lockstep with the high-frequency modes, but in cycle 23 there was
little, if any, change in the low-frequency modes. This marked difference
between the behavior of the low-frequency modes with respect to that of the
high frequency modes allows us to determine
where the changes occurred in the Sun during cycle 23 that was different from changes during
cycle 22 and perhaps even cycle 21.

\begin{figure}
\epsscale{1.0}
\plotone{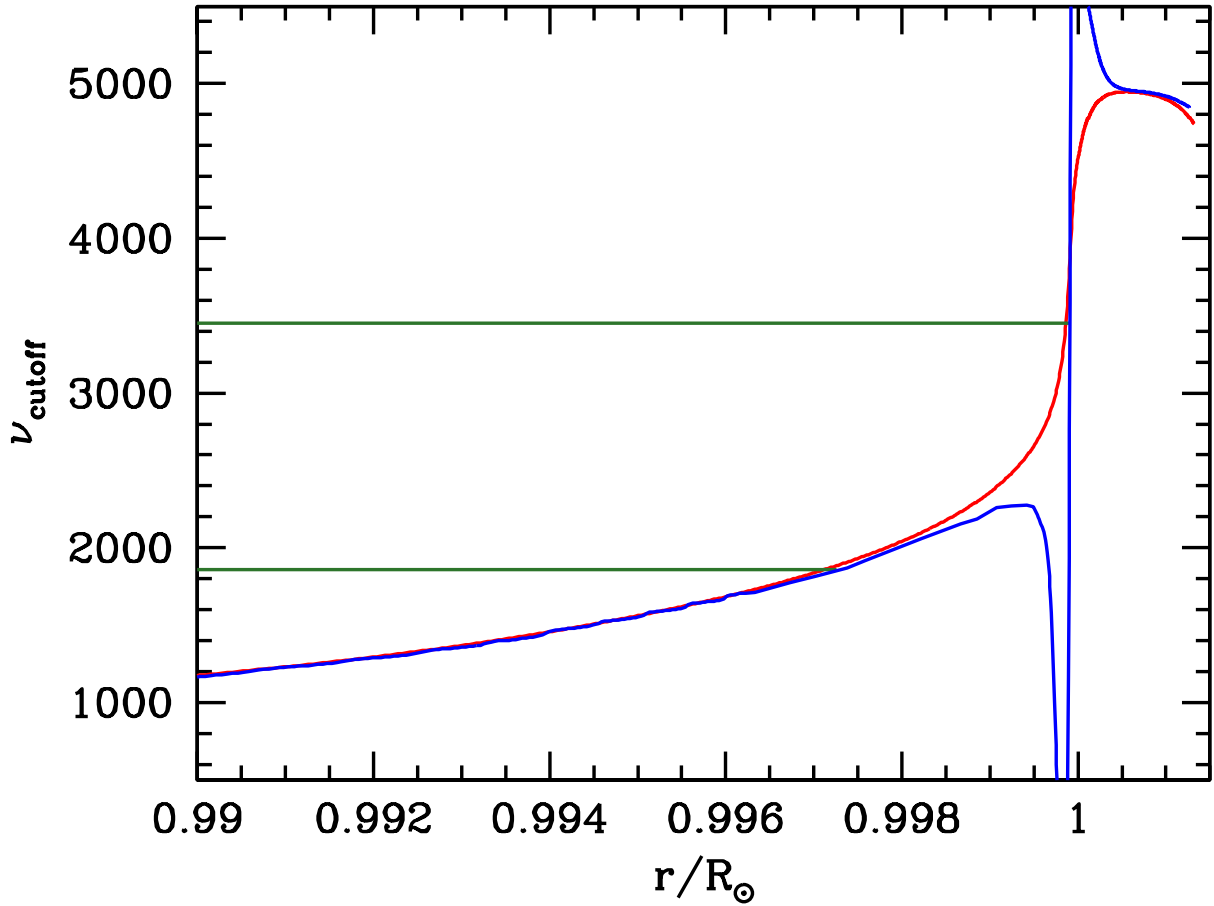}
\caption{
The acoustic cut-off frequency (blue) that approximately defines the upper turning point
of oscillation frequencies of a standard solar model plotted as a function of radius. The curve in
red is the same quantity calculated in the isothermal limit. The green lines represent frequencies at
1860$\mu$Hz and 3450$\mu$Hz. Note that the lower frequency mode is reflected deeper inside the
Sun.}
\label{fig:ut}
\end{figure}

Modes with different
frequencies are reflected back into the Sun at different heights in the atmosphere.
The layer at which a mode is reflected back is usually referred to as the the upper turning point
and this happens at the radius at which the acoustic cut-off frequency of a star equals
the frequency of the mode, i.e., where $\nu_{\rm ac} = \nu$ (see Christensen-Dalsgaard 2003).
Modes have very little sensitivity to the layers beyond this radius.
We show the acoustic cut-off frequency of a standard solar model in Fig.~\ref{fig:ut}.
As can be seen from  this figure, the low-frequency modes that are relatively unchanged during cycle 23 do not penetrate
above about $0.997$R$_\odot$. 
The high- and medium-frequency modes turn closer to the surface. Since the high- and medium-frequency modes
showed changes during both cycle 22 and cycle 23, while the low-frequency modes changed only during
cycle 22, we can conclude that the layer of the Sun that was responsible for the changes in behavior during
cycle 23 more restricted than in cycle 22 and was limited to  the layer above the upper turning point of the low-frequency modes.

\begin{figure}
\plotone{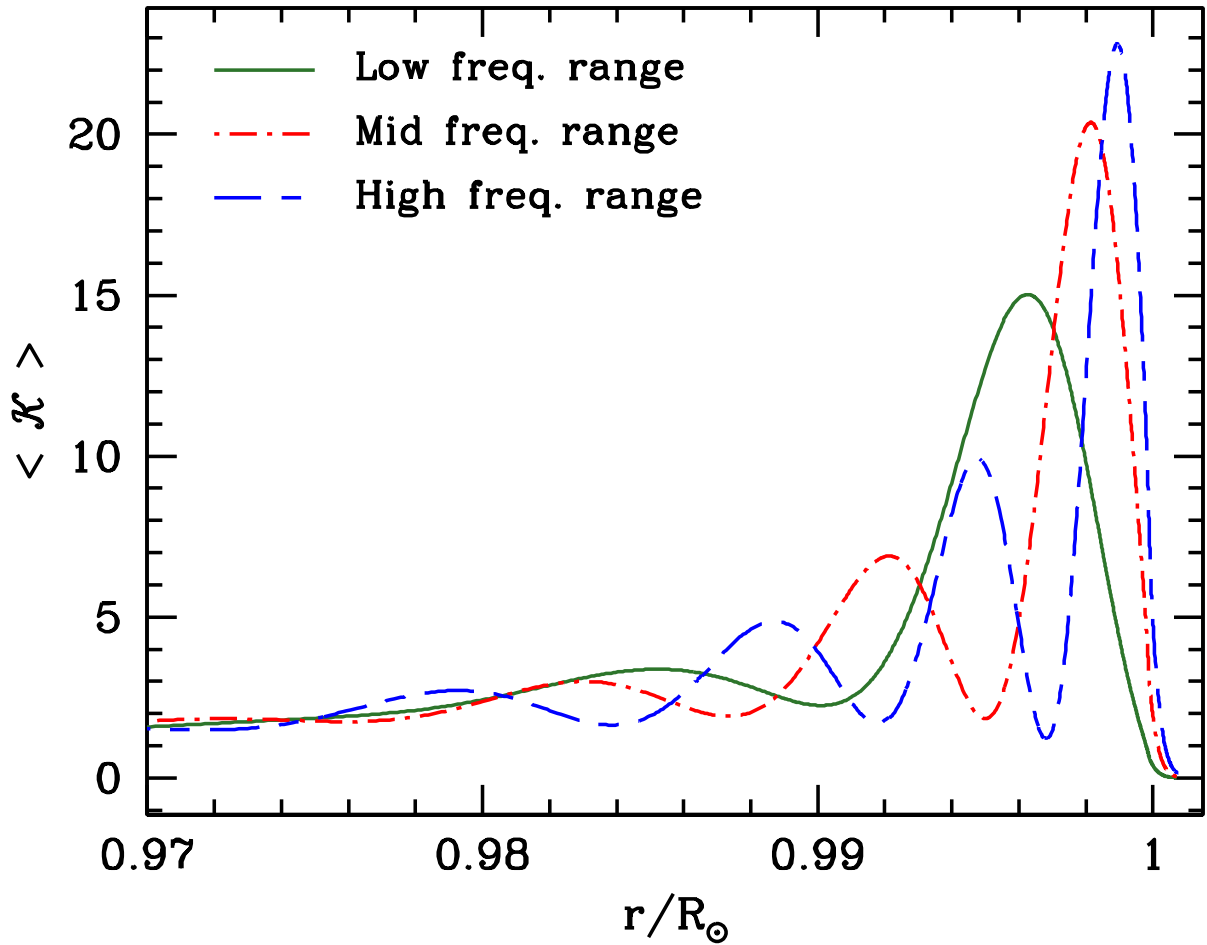}
\caption{
Averaged kernels linking structural changes to frequency changes plotted as a function of solar radius. The
functions correspond to averages over the same modes that were used to calculate the
frequency shifts shown in Figs. 1 and 2}
\label{fig:ker}
\end{figure}

Fig.~\ref{fig:ut} is not the only reason why we think that the differences in the different frequency ranges
point to differences between cycles 22 and 23. Solar frequency differences are related to changes in
solar structure, and can be expressed as:
\begin{equation}
{\delta\nu_i\over\nu_i}=\int {\cal K}_i^{c^2,\rho}(r){\delta c^2\over c^2}(r)dr
+ \int {\cal K}_i^{\rho, c^2}(r){\delta\rho\over\rho}(r)dr,
\label{eq:inv}
\end{equation}
where, ${\cal K}_i^{c^2,\rho}(r)$ is the sound-speed kernel for that mode and ${\cal K}_i^{\rho, c^2}$ is the density
kernel. The kernels  are known functions of solar models (see e.g., Antia \& Basu 1994) and relate the change
in structure to the change in frequency.  The kernels  can be calculated once the
structure and displacement eigenfunctions of a model are known.
The term $\delta c^2/c^2$ is the perturbation in squared sound-speed and $\delta\rho/\rho$ the perturbation in density.
The kernels show the sensitivity of a mode frequency to different parts of the Sun. The larger the amplitude of the
kernels, the most sensitive the mode is to that part of the Sun.
In Fig.~\ref{fig:ker} we show the sound-speed kernels for a solar model.  We averaged the
kernels in the same manner as the modes had been averaged, and these averages {
are shown in the figure for one epoch}
during the cycle 23 descending phase.
As can be seen from Fig.~\ref{fig:ker}, the averaged low-degree modes have maximum sensitivity at 0.9963R$_\odot$, and
the sensitivity falls rapidly at larger radii. The kernels for the mid- and high-frequency ranges peak at
0.9981R$_\odot$ and  0.9989R$_\odot$ respectively. At those radii, the low-frequency kernels have lost most sensitivity.
Thus changes in the Sun that affect both mid- and high-frequency modes can leave the low-frequency modes
almost unaffected.
The fact that the high- and mid-frequency modes change in cycle 23, but the low-frequency modes do not,
points to changes in the Sun that are confined to a layer above a radius of about 0.9965$R_\odot$ i.e., not much
 deeper than about 2400 Km.  In cycle 22, on the other hand, 
changes must have been deeper, otherwise the low-frequency modes would not have
changed to the extent that they had. 
Thus the sub-surface layer of the Sun that shows
the effect of solar magnetic fields was restricted to a thinner layer during most of cycle 23.
A more precise localization of the layer at which the changes occur would require inverting
high-degree ($\ell \ga 1000$) modes that are not available for cycle 22 and not readily available for
cycle 23 either.

There is a small chance that there could have been somewhat deeper changes during cycle 23, but those
would have to have been localized over a very narrow region, centered at about 0.99R$_\odot$
(a depth of 7000 km). While the suggestion of  change at such a specific region sounds somewhat contrived,
it should be noted that it is approximately at this depth and above that the
sub-surface difference in the speed of sound between active regions (containing
strong magnetic fields) and quiet regions changes sign (Basu et al. 2004; Bogart et al. 2008, Moradi et al. 2010).
Given that the reasons for changes in the Sun
 over the solar cycle, as well as structural differences between solar active and quiet regions, lie in
 magnetic fields, it is perhaps not surprisingly that we find evidence of changes at similar depths.

\section{Conclusions}

{
The minimum preceding solar cycle 24 was unusual in its depth and duration.
What causes such deep minima is still a matter of debate. We have analyzed
30 years of helioseismic data collected by the Birmingham Solar Oscillation
Network to compare the solar cycle 23 with cycle 22 in order to determine
whether cycle 23 itself showed peculiarities before the onset of the
peculiar minimum.

We compared the solar-cycle related frequency shifts with two activity
proxies --- the 10.7 cm flux and the international sunspot number --- to find  
that while the activity proxies matched the frequency shifts during cycle
22, this was not the case during cycle 23. In the high- and
medium-frequency ranges the proxies matched the frequency shifts in the  
ascending phase of the solar cycle, but started showing a marked mismatch
just before the cycle 23 maximum and that mismatch continued in the
declining phase of the cycle. In the low-frequency band, the proxies did
not match the cycle-23 shifts at all. This indicates that the peculiarity
in the solar cycle started long before the minimum.

Further analysis of the frequency shifts showed that the high- and
mid-frequency modes in cycles 22 and 23 changed in a similar manner, but
the low-frequency modes did not change much in cycle 23. Thus the Sun had
started showing deviations from ``normal'' activity-related behavior long
before the onset of the deep, ``peculiar'' minimum. Helioseismic signatures
were already present during cycle 23 which indicated that the Sun was
behaving in a way that had not been observed before. With the benefit of
hindsight, these signatures could have been used to forecast the unusual
minimum.

The nature of the changes in solar frequencies in cycle 22 compared with
that of cycle 23 suggests that solar-cycle related changes during cycle 22
occurred deeper in the Sun than in solar cycle 23. Effectively, the layer
of magnetic field  had become thinner after cycle 22 and 
was confined to shallower layers of the Sun.  By using 
well-established helioseismic techniques we are able to localize
these changes. We note also, that the frequency shifts seen in the
low-frequency band are not at the low level seen at previous minima. We can
perhaps conjecture that there is some trapped flux that is unable to make
its way to the surface. In all, there is now the potential for further   
quantitative investigation into the behavior of the magnetic cycle of the
Sun.
}

\acknowledgments This paper utilizes data collected by the Birmingham Solar-Oscillations
Network (BiSON), which is funded by the UK Science Technology and Facilities Council
(STFC). We thank the members of the BiSON team, colleagues at our host institutes, and all
others, past and present, who have been associated with BiSON. SB would like to thank the
HiROS group at the University of Birmingham, U.K., for their hospitality during the period
when this work was conceived and developed. This work is supported by NASA grant
NNX10AE60G and NSF grant ATM-0737770 to SB.

\clearpage

\end{document}